\documentclass[aps,twocolumn,superscriptaddress,preprintnumbers,showpacs]{revtex4}
\usepackage[colorlinks,linkcolor=blue,urlcolor=blue,citecolor=blue,bookmarks,bookmarksnumbered]{hyperref}
\usepackage{bm}
\usepackage{dcolumn}
\usepackage{graphicx}
\usepackage{booktabs}
\usepackage{multirow}
\usepackage{setspace}

\newcommand{\e}{{\mathrm{e}}}
\renewcommand{\i}{{\mathrm{i}}}
\renewcommand{\deg}{^\circ}

\begin{document}

\newcommand*{\PKU}{School of Physics and State Key Laboratory of Nuclear Physics and
Technology, Peking University, Beijing 100871,
China}\affiliation{\PKU}
\newcommand*{\CHEP}{Center for High Energy Physics, Peking University, Beijing 100871, China}\affiliation{\CHEP}

\title{New mixing pattern for neutrinos}

\author{Huilin Qu}\affiliation{\PKU}
\author{Bo-Qiang Ma}\email{mabq@pku.edu.cn}\affiliation{\PKU}\affiliation{\CHEP}

\begin{abstract}
We propose a new mixing pattern for neutrinos with a nonzero mixing
angle $\theta_{13}$. Under a simple form, it agrees well with
current neutrino oscillation data and displays a number of
intriguing features including the $\mu$-$\tau$ interchange symmetry
$|U_{\mu i}|=|U_{\tau i}|$, $(i=1,2,3)$, the trimaximal mixing
$|U_{\e 2}|=|U_{\mu 2}|=|U_{\tau 2}|=1/\sqrt{3}$, the
self-complementarity relation $\theta_1+\theta_3=45\deg$, 
together with the maximal Dirac CP violation as a prediction.

\end{abstract}


\maketitle

After various oscillation experiments for decades, it has been firmly established that neutrinos can transit from one flavor to another in flight due to their mixing. In the standard model of particle physics, the mixing of neutrinos is well described by the Pontecorvo-Maki-Nakagawa-Sakata~(PMNS) matrix~\cite{PMNS}, which is a unitary matrix connecting neutrino flavor eigenstates and mass eigenstates. The PMNS matrix is conventionally expressed in the standard parametrization, i.e., the Chau-Keung (CK) scheme~\cite{CK}, by three angles
$\theta_{12}$, $\theta_{13}$, $\theta_{23}$ and one CP-violating phase angle $\delta$ in the form
\begin{widetext}
\begin{eqnarray}
U_{\rm CK} = \left (
\begin{array}{ccc}
c^{~}_{12} c_{13}         & s^{~}_{12} c_{13}       & s_{13}
\e^{-\i\delta}\cr -c^{~}_{12} s_{23} s_{13} \e^{\i\delta}- s^{~}_{12}
c_{23} & -s^{~}_{12} s_{23} s_{13}\e^{\i\delta} + c^{~}_{12} c_{23}
& s_{23} c_{13} \cr -c^{~}_{12} c_{23} s_{13} \e^{\i\delta}+
s^{~}_{12} s_{23} & -s^{~}_{12} c_{23} s_{13} \e^{\i\delta}-
c^{~}_{12} s_{23} & c_{23} c_{13}
\end{array}
\right),
\label{eq:CK}
\end{eqnarray}
\end{widetext}
where $s_{ij}={\rm sin}\theta_{ij}$ and $c_{ij}={\rm
cos}\theta_{ij}$ ($i,j=1,2,3$), and an additional factor $P_\nu={\rm
Diag}\{\e^{-\i\alpha/2},\e^{-\i\beta/2},1\}$ should be multiplied to the right if neutrinos are Majorana particles.

Different from the Cabibbo-Kobayashi-Maskawa (CKM)
matrix~\cite{CKM1,CKM2} for quark mixing, where mixing angles are
small and the CKM matrix is close to the identity matrix \cite{PDG},
the mixing angles for neutrinos are much larger, and the PMNS matrix
exhibits a significant deviation from the identity matrix. Thus, a
number of simple mixing patterns with finite mixing angles were
proposed and extensively studied, such as the bimaximal~(BM) mixing
pattern~\cite{bi}
\begin{eqnarray}
U_{\rm BM}=\left(
\begin{array}{ccc}
\frac{1}{\sqrt{2}} & \frac{1}{\sqrt{2}}  & 0 \\
-\frac{1}{2} & \frac{1}{2} & \frac{1}{\sqrt{2}}  \\
\frac{1}{2} & -\frac{1}{2} & \frac{1}{\sqrt{2}}
\end{array}\right), \ \
\label{eq:BM}
\end{eqnarray}
with $\theta_{12}=\theta_{23}=45^\circ$, and the tribimaximal~(TB) mixing
pattern~\cite{tri}
\begin{eqnarray}
U_{\rm TB}=\left(
\begin{array}{ccc}
\sqrt{\frac{2}{3}}& \frac{1}{\sqrt{3}} & 0 \\
-\frac{1}{\sqrt{6}} & \frac{1}{\sqrt{3}} & \frac{1}{\sqrt{2}} \\
\frac{1}{\sqrt{6}} &  -\frac{1}{\sqrt{3}}& \frac{1}{\sqrt{2}}
\end{array}\right),
\label{eq:TB}
\end{eqnarray}
with $\theta_{12}=35.26^\circ$ and $\theta_{23}=45^\circ$. However, in both cases the smallest mixing angle $\theta_{13}$ vanishes, which is incompatible with a nonzero and relatively large $\theta_{13}$ established by recent accelerator and reactor neutrino oscillation experiments~\cite{expt,daya-bay,Ahn:2012nd}.
There have been attempts to build a new mixing pattern with a nonzero $\theta_{13}$~\cite{Zheng:2011uz} based on a self-complementary relation~\cite{Zhang:2012xu} $\theta_{12} + \theta_{13}
=\theta_{23}= 45^\circ$ between neutrino mixing angles, yet the resulting mixing matrix is far from simplicity. A new mixing pattern with a sizable $\theta_{13}$ and a simple form at the same time is being called for.

In this paper we propose a new mixing pattern of neutrinos with a nonzero $\theta_{13}$. It is both simple in form and close to current neutrino data. In addition, it displays a number of phenomenological relations including the $\mu$-$\tau$ interchange symmetry, the trimaximal mixing and the self-complementarity. The maximal Dirac CP violation is also predicted in this mixing pattern.

\begin{figure}
\centering
\includegraphics[width=\linewidth]{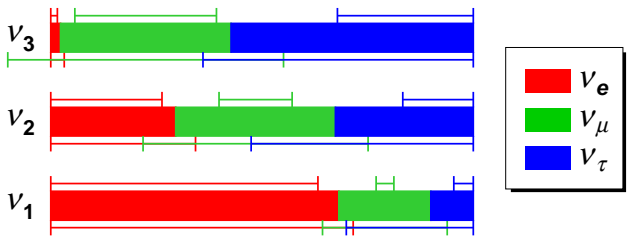}
\caption{The mixing spectrum of neutrinos.  $\nu_1,\,\nu_2,\,\nu_3$ are mass eigenstates with masses $m_1,\,m_2,\,m_3$, respectively, and $\nu_e,\,\nu_\mu,\,\nu_\tau$ are flavor eigenstates. The length of each colored bar is proportional to the modulus squared of the mixing matrix element, $|U_{\alpha i}|^2$, calculated from the best-fit values of neutrino mixing parameters in Ref.~\cite{globalfit}. The line segment plotted above (below) each colored bar represents the $3\sigma$ lower (upper) bound of the corresponding $|U_{\alpha i}|^2$. }
\label{fig:plot}
\end{figure}

In the search of a new mixing pattern for neutrinos, it is important to inspect the current neutrino oscillation data and see where we stand. Fig.~\ref{fig:plot} shows the mixing spectrum of neutrinos, plotted according to a global fit result of neutrino oscillation data \cite{globalfit}. We denote neutrino mass eigenstates with the mass $m_i$ by $\nu_i$ ($i=1,2,3$) and flavor eigenstates by $\nu_\alpha$ ($\alpha=e,\mu,\tau$). Each colored bar represents the modulus squared of the mixing matrix element, $|U_{\alpha i}|^2$. Here some features should be noticed:
\begin{enumerate}
\item While the proportion of $\nu_e$ in $\nu_3$ is quite small, it is not negligible, i.e., $|U_{e3}|\neq 0$.
\item The mass eigenstate $\nu_2$ is almost equally shared by $\nu_{e},\,\nu_{\mu}$ and $\nu_{\tau}$, which is usually referred to as ``trimaximal mixing''.
\item Although it is not perfectly satisfied, the long-studied $\mu$-$\tau$ interchange symmetry \cite{mutau1,mutau2}, i.e., $|U_{\mu i}|=|U_{\tau i}|,\ (i=1,2,3)$, still holds approximately considering the experimental uncertainties and the undetermined CP-violating phase.
\end{enumerate}
When we are looking for a new mixing pattern, it is necessary to take these features into account.

Besides, another interesting phenomenological relation, the self-complementarity relation~\cite{Zheng:2011uz, Zhang:2012xu}, also catches our eyes. Unlike the above properties that are stated at the matrix element level, the self-complementarity can only be studied after we choose a specific parametrization of the mixing matrix. Originally, it is observed that mixing angles in the standard CK scheme are in accord with the relation $\theta_{12} + \theta_{13}=45^\circ$ \cite{Zheng:2011uz}.
However, similar to the quark-lepton complementarity (QLC) relation~\cite{QLC1,phenomenology,Zhang:2012zh} which is in fact parametrization-dependent~\cite{Jarlskog:2005jn}, the validity of the self-complementarity relation also varies significantly in different parametrizations~\cite{Zhang:2012}. From Ref.~\cite{Zhang:2012} we find that, among the nine different schemes to parametrize the PMNS matrix, the self-complementarity is best satisfied not in the standard CK scheme, but in a different parametrization denoted by P4 in that reference. This motivates us to consider the self-complementarity in this new parameterization, which is of the form
\begin{widetext}
\begin{eqnarray}
U(\theta_1,\theta_2,\theta_3,\phi)=
\left(\begin{array}{ccc}
            c_1c_3 & \hphantom{-}s_1 & -c_1s_3\\
            -s_1c_2c_3+s_2s_3\e^{-\i\phi} & \hphantom{-}c_1c_2 & \hphantom{-} s_1c_2s_3+s_2c_3\e^{-\i\phi}\\
            \hphantom{-}s_1s_2c_3+c_2s_3\e^{-\i\phi} & -c_1s_2 & -s_1s_2s_3+c_2c_3\e^{-\i\phi}
      \end{array}\right),
      \label{eq:par}
\end{eqnarray}
\end{widetext}
where $s_{i}={\rm sin}\theta_{i}$ and $c_{i}={\rm
cos}\theta_{i}$ ($i=1,2,3$), and the CP-violating phase is denoted by $\phi$ so as to distinguish it from the CP-violating phase $\delta$ in the standard CK scheme. The self-complementarity in this parametrization is defined as $\theta_1+\theta_3=45\deg$.

Working in this parametrization, we seek a particular mixing pattern in which the above three features and the self-complementarity hold exactly, written explicitly as
\begin{enumerate}
\item a nonzero $|U_{e3}|$;
\item the $\mu$-$\tau$ interchange symmetry in modulus, i.e., $|U_{\mu i}|=|U_{\tau i}|,\ (i=1,2,3)$;
\item the trimaximal mixing in $\nu_2$, i.e., $|U_{\e 2}|=|U_{\mu 2}|=|U_{\tau 2}|=\frac{1}{\sqrt{3}}$;
\item the self-complementarity relation, i.e., $\theta_1+\theta_3=45\deg$.
\end{enumerate}
Among these four requirements, the only viable solution for the first and second ones is $\theta_2=45\deg$ and $\phi=\pm 90\deg$, and the third and fourth ones give rise to $\sin\theta_1=\frac{1}{\sqrt{3}}$ and $\sin\theta_3=\sin(45\deg-\theta_1)=\frac{1}{\sqrt{3}}-\frac{1}{\sqrt{6}}$, respectively. Substituting these values into Eq.~(\ref{eq:par}), we obtain the mixing matrix satisfying all the four requirements:
\begin{eqnarray}
U_0=\left(
\begin{array}{ccc}
 \hphantom{0}\frac{\sqrt{2}+1}{3}  & \hphantom{-}\frac{1}{\sqrt{3}} & -\frac{\sqrt{2}-1}{3}  \\[1.5mm]
 -\frac{\sqrt{2}+1}{6} \mp \i \frac{\sqrt{6}-\sqrt{3}}{6} & \hphantom{-}\frac{1}{\sqrt{3}} & \hphantom{-}\frac{\sqrt{2}-1}{6} \mp \i \frac{\sqrt{6}+\sqrt{3}}{6} \\[1.5mm]
 \hphantom{-}\frac{\sqrt{2}+1}{6} \mp \i \frac{\sqrt{6}-\sqrt{3}}{6} & -\frac{1}{\sqrt{3}} & -\frac{\sqrt{2}-1}{6} \mp \i \frac{\sqrt{6}+\sqrt{3}}{6}
\end{array}
\right).
\label{eq:U0}
\end{eqnarray}
However, since the phase convention varies in different parametrizations, it would be more useful to write down the moduli of the mixing matrix, which is invariant under reparametrization,
\begin{eqnarray}
\left|U_0\right|=\left(
\begin{array}{ccc}
 \frac{\sqrt{2}+1}{3}  & \frac{1}{\sqrt{3}} & \frac{\sqrt{2}-1}{3}  \\[1.5mm]
 \frac{\sqrt{3-\sqrt{2}}}{3} & \frac{1}{\sqrt{3}} & \frac{\sqrt{3+\sqrt{2}}}{3} \\[1.5mm]
 \frac{\sqrt{3-\sqrt{2}}}{3} & \frac{1}{\sqrt{3}} & \frac{\sqrt{3+\sqrt{2}}}{3}
\end{array}
\right).
\label{eq:pattern}
\end{eqnarray}
Eq.~(\ref{eq:pattern}) is our proposal for a new mixing pattern of
neutrinos. It is independent of the parametrization we choose. We
see that this new mixing pattern is strikingly simple and elegant,
with only the smallest positive integers 1, 2, and 3 appearing in
the mixing pattern. Yet it satisfies all the phenomenological
relations as stated above.

In order to compare this new mixing pattern with neutrino oscillation data, we first solve for the mixing angles in the standard parametrization by equating the moduli of Eq.~(\ref{eq:CK}) with corresponding elements in Eq.~(\ref{eq:pattern}). After some simple and straightforward calculations, we obtain
\begin{eqnarray}
\begin{array}{ccl}
\sin\theta_{12}&=&\sqrt{\frac{3}{2 \left(3+\sqrt{2}\right)}},\\[2mm] \sin\theta_{13}&=&\frac{\sqrt{2}-1}{3},\\[2mm]
\sin\theta_{23}&=&\frac{1}{\sqrt{2}},\\[2mm]
\cos\delta&=&0,
\end{array}
\label{eq:sin0}
\end{eqnarray}
or
\begin{eqnarray}
\begin{array}{ccl}
\theta_{12}&\simeq& 35.66^\circ,\\
\theta_{13}&\simeq& 7.94^\circ,\\
\theta_{23}&=&45^\circ,\\
\delta&=&\pm 90^\circ.
\end{array}
\label{eq:angles0}
\end{eqnarray}
The recent global fit results \cite{globalfit} are listed in Table \ref{tab:fit}. In Ref. \cite{globalfit} two fits based on different assumptions about the reactor fluxes are provided, and since their values vary only slightly, only the ``Free Fluxes + RSBL'' case is quoted here.
\begin{table}
      \caption{Results for the neutrino mixing angles and the Dirac CP-violating phase taken from the global fit to neutrino oscillation data \cite{globalfit}. In Ref. \cite{globalfit} two fits based on different assumptions about the reactor fluxes are provided, and only the ``Free Fluxes + RSBL'' case is listed here. }
      \label{tab:fit}
      \begin{ruledtabular}
      \begin{tabular}{ccc}
      Parameter & Best fit$\pm 1\sigma$ & $3\sigma$ range\\\hline
      \noalign{\vspace{0.5ex}}
      $\theta_{12}/\deg$ & $33.36_{-0.78}^{+0.81}$ & $31.09 \to 35.89$\\
      \noalign{\vspace{0.5ex}}
      $\theta_{13}/\deg$ & $8.66_{-0.46}^{+0.44}$ & $7.19 \to 9.96$\\
      \noalign{\vspace{0.5ex}}
      $\theta_{23}/\deg$ & $40.0_{-1.5}^{+2.1} \oplus 50.4_{-1.3}^{+1.3}$ & $35.8 \to 54.8$ \\
      \noalign{\vspace{0.5ex}}
      $\delta/\deg$ & $300_{-138}^{+66}$ & $\hphantom{0.0}0 \to 360\hphantom{.}$\\
      \end{tabular}
      \end{ruledtabular}
\end{table}
We see that all the parameters in our new mixing pattern are compatible with experimental measurements, lying in the $3\sigma$ range of the global fit result.

It is worthy to note that, when examined in the standard parametrization, our new mixing pattern displays a maximal mixing angle $\theta_{23}=45^\circ$ and a maximal Dirac CP-violating phase $\delta=\pm 90^\circ$. The maximal Dirac CP violation $\delta=\pm 90^\circ$ is also consistent with a previous phenomenological analysis result $\delta=(85.39^{+4.76}_{-1.82})\deg$, derived from the hypothesis that CP violation is maximal in the Kobayashi-Maskawa (KM) scheme \cite{Zhang:2012ys}. Although the strict self-complementarity relation is broken by $\theta_{12}+\theta_{13}\simeq 43.6\deg\neq 45\deg$, it is rather close to $45\deg$, which is in accord with the original discovery of the self-complementarity in the standard parametrization \cite{Zheng:2011uz}.

It is also worthy to note that, our new mixing pattern, specifically Eq.~(\ref{eq:U0}), can actually be viewed as the particular case of the tri-$\chi$maximal mixing~\cite{Harrison:2002kp}, in which $\sin\chi=\frac{\sqrt{2}-1}{\sqrt{6}}$. At the same time, the Jarlskog CP-violating invariant~\cite{jarlskog} $|J|=\frac{1}{18 \sqrt{3}}$, which attains exactly $1/3$ of its maximum value of the tri-$\chi$maximal mixing.

Since the Dirac CP-violating phase $\delta$ is so far not determined by any experiment, it would be helpful to write down the mixing matrix in the standard parametrization with $\delta$ left as a free parameter, which can be further probed by the upcoming experiments. Substituting the three mixing angles in Eq.~(\ref{eq:sin0}) into the standard parametrization Eq.~(\ref{eq:CK}), we get
\begin{equation}
U=\left(
\begin{array}{ccc}
 \hphantom{-}\frac{1+\sqrt{2}}{3} & \frac{1}{\sqrt{3}} & \frac{-1+\sqrt{2}}{3}  \e^{-\i \delta } \\[1.5mm]
 -\frac{3 \sqrt{3}+\e^{\i \delta }}{6 \sqrt{3+\sqrt{2}}} & \frac{\hphantom{-}3 \left(1+\sqrt{2}\right)-\left(\sqrt{6}-\sqrt{3}\right) \e^{\i \delta }}{6 \sqrt{3+\sqrt{2}}} & \frac{\sqrt{3+\sqrt{2}}}{3} \\[1.5mm]
 \hphantom{-}\frac{3 \sqrt{3}-\e^{\i \delta }}{6 \sqrt{3+\sqrt{2}}} & \frac{-3 \left(1+\sqrt{2}\right)-\left(\sqrt{6}-\sqrt{3}\right) \e^{\i \delta }}{6 \sqrt{3+\sqrt{2}}} & \frac{\sqrt{3+\sqrt{2}}}{3}
\end{array}\right).
\label{eq:matrix_CP}
\end{equation}
In the case of $\delta=\pm 90\deg$, the moduli of Eq.~(\ref{eq:matrix_CP}) will take the form of our new mixing pattern Eq.~(\ref{eq:pattern}).


In conclusion, we propose a new mixing pattern for neutrinos, as
shown in Eq.~(\ref{eq:pattern}), which agrees well with current
neutrino oscillation data, especially the nonzero and relatively
large value of $\theta_{13}$. While extremely simple in form, this
new mixing pattern demonstrates a series of intriguing features
including the $\mu$-$\tau$ interchange symmetry in modulus, the
trimaximal mixing, the self-complementarity relation, 
together with a natural prediction of the maximal Dirac CP violation.
The new mixing pattern may imply certain
family symmetry, which would help us unravel the mystery of masses
and mixing structures of fermions.

\begin{acknowledgments}
This work is supported by National Natural Science Foundation of China (Grants No.~11021092, No.~10975003, No.~11035003, and No.~11120101004) and the National Fund for Fostering Talents of Basic Science (Grant No.~J1103205). It is also supported by the Undergraduate Research Fund of Education Foundation of Peking University. 
\end{acknowledgments}

\end{document}